\documentclass[aps,pra,reprint,superscriptaddress,showkeys,amsmath,amssymb]{revtex4-2}
\usepackage[utf8]{inputenc}
\usepackage{graphicx}
\usepackage[hidelinks]{hyperref}

\makeatletter
\def\maketitle{
\@author@finish
\title@column\titleblock@produce
\suppressfloats[t]
}%
\makeatother

\newcommand{\figref}[1]{\textbf{Figure~\ref{#1}}}
\newcommand{\eqrefrb}[1]{\textbf{Equation~\ref{#1}}}

\makeatletter
\renewcommand{\fnum@figure}{\textbf{Figure~\thefigure}}
\makeatother

\begin{document}


\author{R. Barczyk}
\affiliation{Center for Nanophotonics, AMOLF, Science Park 104, 1098 XG Amsterdam, The Netherlands}
\author{N. Parappurath}
\affiliation{Center for Nanophotonics, AMOLF, Science Park 104, 1098 XG Amsterdam, The Netherlands}
\author{S. Arora}
\affiliation{Kavli Institute of Nanoscience, Delft University of Technology, 2600 GA, Delft, The Netherlands}
\author{T. Bauer}
\affiliation{Kavli Institute of Nanoscience, Delft University of Technology, 2600 GA, Delft, The Netherlands}
\author{L. Kuipers}
\affiliation{Kavli Institute of Nanoscience, Delft University of Technology, 2600 GA, Delft, The Netherlands}
\author{E. Verhagen}\email{e.verhagen@amolf.nl}
\affiliation{Center for Nanophotonics, AMOLF, Science Park 104, 1098 XG Amsterdam, The Netherlands}

\title{Interplay of leakage radiation and protection in topological photonic crystal cavities}

\begin{abstract}

The introduction of topological concepts to the design of photonic crystal cavities holds great promise for applications in integrated photonics due to the prospect of topological protection. This study examines the signatures of topological light confinement in the leakage radiation of two-dimensional topological photonic crystal cavities. The cavities are implemented in an all-dielectric platform that features the photonic quantum spin Hall effect at telecom wavelengths and supports helical edge states that are weakly coupled to the radiation continuum. The modes of resonators scaling down to single point defects in the surrounding bulk lattice are characterized via spectral position and multipolar nature of the eigenstates. The mode profiles in real and momentum space are mapped using far field imaging and Fourier spectropolarimetry, revealing how certain properties of the cavity modes reflect on their origin in the topological bandstructure. This includes band-inversion-induced confinement and inverted scaling of mode spectra for trivial and topological defect cavities. Furthermore, hallmarks of topological protection in the loss rates are demonstrated, which are largely unaffected by cavity shape and size. The results constitute an important step toward the use of radiative topological cavities for on-chip confinement of light, control of emitted wave fronts, and enhancing light-matter interactions.

\end{abstract}

\maketitle

\begin{figure*}[hbt]
  \centering
  \includegraphics[width=\linewidth]{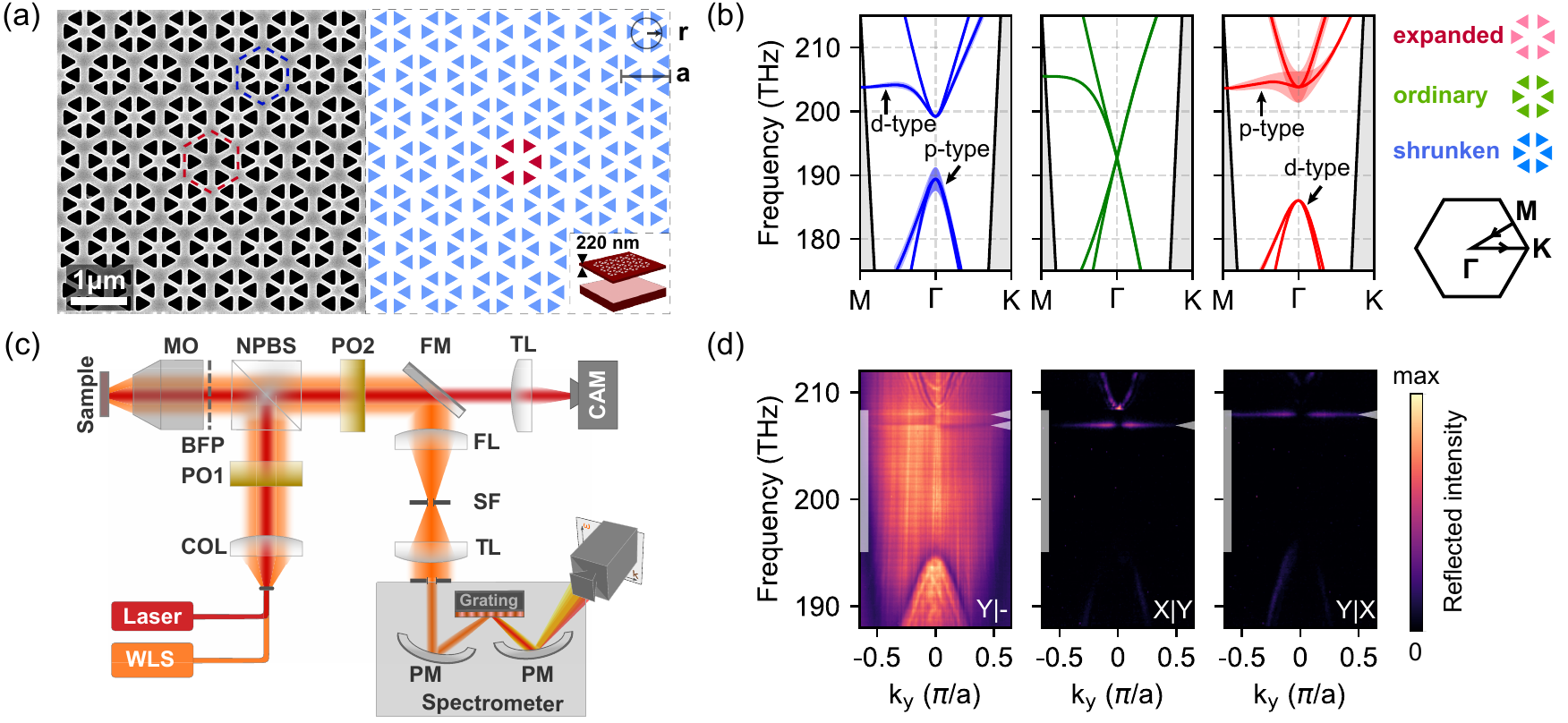}
  \caption{System, setup, and typical measurements. a) Topological photonic crystal cavity in a silicon membrane. Left: scanning electron micrograph of a defect cavity made from a single topologically non-trivial (expanded) unit cell, surrounded by a trivial (shrunken) bulk, with unit cells indicated by red and blue dashed hexagons, respectively. Right: schematic illustration of the system with an inset of the suspended photonic crystal membrane geometry. The distance $r$ of the scattering sites to the unit cell center and the lattice constant $a$  are indicated. b) Numerically obtained photonic bands for an expanded (blue), ordinary $C_6$-symmetric (green), and shrunken (red) bulk. We denote the dipolar (p-type) and quadrupolar (d-type) nature of the bulk bands and indicate the associated linewidths by colored shadings (scaled up $5\times$ for visibility). Regions outside of the light cone are shaded in gray. Inset: computed trajectory within the fundamental Brillouin zone. c) Scheme of the experimental Fourier-spectropolarimetry setup used for angularly resolved measurement of the cavity modes' dispersion (orange path) and far field radiation profile (red path). See the Experimental Section for details and abbreviations. d) Mode spectrum of the system shown in a) for various settings of polarizer and analyzer (polarizer$|$analyzer). Recognizable cavity modes as well as the bulk band gaps are highlighted by white pointers and light shaded bars, respectively.}
  \label{fig:1}
\end{figure*}

\section{Introduction}

Photonic crystal (PhC) resonators represent versatile building blocks for the control and manipulation of light on wavelength scales,\:\cite{joannopoulosPhotonicCrystalsMolding2008} offering highly desirable functionality for on-chip integration such as sensing,\:\cite{pitruzzelloPhotonicCrystalResonances2018} delay lines,\:\cite{hafeziRobustOpticalDelay2011a} and quantum information processing.\:\cite{dietrichGaAsIntegratedQuantum2016} Since the confinement of light to small mode volumes renders resonators more prone to radiative losses, i.e., scattering and out-of-plane leakage, extensive effort has been conducted to realize miniaturized whispering-gallery-mode (WGM) and lattice-defect PhC cavities with various quality factors and mode volumes through optimization of design down to the nanometer level.\:\cite{srinivasanMomentumSpaceDesign2002, akahaneHighPhotonicNanocavity2003, akahaneFinetunedHighQPhotoniccrystal2005, foremanWhisperingGalleryMode2015} In this context, the concept of topological insulators provides an interesting new paradigm for photonic design, including that of PhC resonators, as it offers inherent robustness to specific forms of imperfections.\:\cite{hasan_colloquium_2010, RevModPhys.83.1057, asboth2016short}\\
Topological insulators are characterized by a non-trivial topology of the material's bandstructure, originating from the geometrical structure of the associated wave functions in momentum space. While being insulating in the bulk, the boundary of a topological insulator supports propagating edge states,\:\cite{renTopologicalPhasesTwodimensional2016a, asboth2016short} constituting a transmission line resilient to backscattering by a wide range of perturbations which respect the symmetry that underlies the topological protection. The transfer of topological ideas from the electronic to the photonic domain lead to the advent of photonic topological insulators (PTIs) with unprecedented light-guiding capabilities.
Apart from Chern-type PTIs that break time-reversal-symmetry,\:\cite{haldanePossibleRealizationDirectional2008}\;time-reversal-invariant\;PTIs that rely on certain spatial symmetries for a protection mechanism were proposed and successfully implemented in passive dielectric PhCs.\:\cite{fuTopologicalCrystallineInsulators2011, PhysRevLett.114.223901, Barik_2016, andersonUnidirectionalEdgeStates2017,barikTopologicalQuantumOptics2018} Due to a lack of fermionic Kramer's doubling in bosonic systems, PTIs that mimic the quantum valley-Hall effect (QVHE)\:\cite{aroraDirectQuantificationTopological2021, maAllSiValleyHallPhotonic2016, dongValleyPhotonicCrystals2017} or quantum spin-Hall effect (QSHE)\:\cite{parappurathDirectObservationTopological2020, andersonUnidirectionalEdgeStates2017, barikTopologicalQuantumOptics2018} take a geometrical approach to construct pseudospins and corresponding spin-momentum locking,\:\cite{bliokhQuantumSpinHall2015,salaSpinOrbitCouplingPhotons2015} a luring proposition for chiral quantum optics.\:\cite{lodahlChiralQuantumOptics2017, barikTopologicalQuantumOptics2018}
Building upon the light-guiding capabilities of PTIs, one can design ring-like PhC cavities based on the photonic QVHE\:\cite{barikChiralQuantumOptics2020, jiRobustFanoResonance2021, kimMultibandPhotonicTopological2021, mehrabadChiralTopologicalPhotonics2020, zengElectricallyPumpedTopological2020a} and QSHE.\:\cite{sirokiTopologicalPhotonicsCrystals2017, yangTopologicalWhisperingGallery2018, gaoDiracvortexTopologicalCavities2020, jalalimehrabadSemiconductorTopologicalPhotonic2020, sunTopologicalRingcavityLaser2021, hafeziImagingTopologicalEdge2013} While QVHE-type PTIs operate below the light line, edge states in QSHE-type PTIs are intrinsically leaky,\:\cite{gorlachFarfieldProbingLeaky2018} providing a direct near-to-far field interface. Control over far field radiation properties of PhC cavities is highly relevant for applications relying on efficient emission (photoluminescence, lasing) or in-coupling (optical pumping, add-drop functionality) of light from and into nanophotonic devices, respectively. Various studies on conventional PhC cavities are therefore dedicated to the optimization of far field coupling for e.g. light-matter interfaces, sensors, and surface-emitting lasers.\:\cite{jagerskaRefractiveIndexSensing2010, hamelSpontaneousMirrorsymmetryBreaking2015, iwahashiHigherorderVectorBeams2011, yangSpinMomentumLockedEdgeMode2020, shaoHighperformanceTopologicalBulk2020b} With its inherent coupling to radiation, QSHE-type cavities could thus provide an interesting platform for such applications. At the same time, as the leakage is not contained in the topological description of the system, intriguing fundamental questions arise: how do the cavities' radiation losses affect topological robustness and, conversely, how does the PhC's band structure control the emitted radiation? The fact that intrinsic loss and topological protection naturally coalesce in QSHE-type PhC cavities calls for an examination on the consequences of the mutual presence of these seemingly opposing features. \\
Here, we present an experimental study of the radiative optical properties of Si nanocavities using a topological PhC platform featuring the photonic QSHE at telecom frequencies. Contrary to earlier theoretical\:\cite{sirokiTopologicalPhotonicsCrystals2017} and experimental studies,\:\cite{yuCriticalCouplingsTopologicalinsulator2021, yangTopologicalWhisperingGallery2018, jalalimehrabadSemiconductorTopologicalPhotonic2020, jalalimehrabadSemiconductorTopologicalPhotonic2020, gaoDiracvortexTopologicalCavities2020, yangSpinMomentumLockedEdgeMode2020, shaoHighperformanceTopologicalBulk2020b} we investigate resonators that are coupled to the radiation continuum and scale down to the size of a single point defect in the surrounding bulk lattice, where the common interpretation of the interface as a ring-like topological waveguide breaks down. We find that the origin of the cavity modes in the topological bandstructure is still manifested in the mode properties. To acquire full knowledge over the cavities' spectral, angular, spatial, and polarimetric characteristics, we employ Fourier spectropolarimetry of the cavity far field radiation. We elucidate the modes' multipolar nature and scaling with cavity size, and identify hallmarks of topological behavior such as band-inversion-induced confinement, inverted mode spectra for shrunken and expanded cavities, and directional emission in vortex beam modes. We demonstrate that, in contrast to regular PhC cavities and WGM resonators, the mode quality factors are largely insensitive to the cavity size, and examine the robustness of mode spectra against strong deformations of the cavity shape.

\section{Results and Discussion}

\subsection{Topological Photonic Crystal Design and Characterization}

Our implementation of a topological PhC platform follows the symmetry-based approach outlined by Wu \& Hu.\:\cite{PhysRevLett.114.223901} The scheme relies on the geometric properties of the PhC which, in our case, is formed by equilateral triangular air holes perforating a silicon membrane.\:\cite{Barik_2016} We fabricate these structures via electron beam lithography and reactive ion and wet etching of a silicon-on-insulator substrate (see the Experimental Section for more details).\:\cite{reardon_fabrication_2012}
\figref{fig:1}a shows a top view scanning electron micrograph of a fabricated nanocavity, next to a pictorial representation illustrating the two lattice types constituting the system. The inset represents a three-dimensional cross-cut of the sample, displaying a suspended PhC membrane. The relation between unit cell geometry and photonic dispersion for TE-polarized bulk crystal modes is presented in Figure \ref{fig:1}b. Considering a six-site unit cell of a perfect triangular lattice, a Dirac-like linear dispersion is found at the Gamma point in the center of the Brillouin zone (Figure \ref{fig:1}b, center panel).\:\cite{PhysRevLett.114.223901} It is four-fold degenerate, as it originates from band folding of the Dirac points in the dispersion of a graphene-like lattice at the K and K' points when considering a fundamental two-site rhombic unit cell. Spatial symmetry breaking creates bandgaps with non-trivial topological nature through the lifting of this Dirac-point degeneracy.\:\cite{asboth2016short} Bandgaps of different topological order are created by continuously deforming the unit cell, radially shifting the holes while preserving $C_6$ symmetry to obtain either the `shrunken' (Figure \ref{fig:1}b, left) or the `expanded' (Figure \ref{fig:1}b, right panel) design. 
The size of the bandgap is an important parameter for resonators as it relates to spatial confinement, i.e., the mode volume.\:\cite{joannopoulosPhotonicCrystalsMolding2008} For our sample geometry (see the Experimental Section for detailed parameters), we numerically obtain relative bandgap sizes of $\Delta\omega/\omega_0 \approx 5.0\,\%$ and $7.4\,\%$ for the shrunken and expanded case, centered around $194.3\,$THz and $196.5\,$THz, respectively. The bandgap size may be tuned by varying lattice parameters such as the expansion factor $r/r_0$, where $r$ describes the radius of the centroids of the triangular air holes, with $r=r_0$ for the ordinary lattice. Larger deviations $|r/r_0-1|$ are associated with smaller mode volumes.\:\cite{jalalimehrabadSemiconductorTopologicalPhotonic2020} Without detailed optimization of the design to minimize the leakage rate through interference,\:\cite{akahaneHighPhotonicNanocavity2003, portalupiPlanarPhotonicCrystal2010, srinivasanMomentumSpaceDesign2002} in regular PhC cavities a smaller mode volume is generally accompanied by larger radiative loss. In our case, $r/r_0=0.91$ and $1.09$ for the shrunken and expanded unit cells, respectively.\:\cite{parappurathDirectObservationTopological2020}
The topological nature of the system is associated with band inversion, like in the one-dimensional Su-Schrieffer-Heeger chain:\:\cite{asboth2016short} a continuous transformation from shrunken to expanded geometry inverts the ratio of intra- to inter-cell coupling of scattering sites, necessarily accompanied by intermediate closing of the bandgap. In consequence, the nature of the states populating the band edges is inverted. While the states of the bottom (top) bulk bands in the shrunken (expanded) lattice possess an out-of-plane magnetic field $H_z$ that resembles dipolar p-orbitals, responsible for the radiative loss, the top (bottom) bands feature sub-radiant quadrupolar d-orbitals.\:\cite{PhysRevLett.114.223901, Barik_2016, gorlachFarfieldProbingLeaky2018, parappurathDirectObservationTopological2020} The difference in leakiness is also reflected in the radiative linewidths (Figure \ref{fig:1}b, colored shading), which is about an order of magnitude larger for p-type bands.\:\cite{gorlachFarfieldProbingLeaky2018, parappurathDirectObservationTopological2020}
Although the true topological character of the platform has been a topic of debate,\:\cite{orazbayevQuantitativeRobustnessAnalysis2019,depazEngineeringFragileTopology2019} recent studies\:\cite{palmerBerryBandsPseudospin2021} support the argument that the bulk topology of the lattices is indeed analogous to a $\mathbb{Z}_2$ topological insulator in the QSHE,\:\cite{asboth2016short} with the expanded lattice being in a non-trivial phase.\:\cite{Barik_2016} Bulk-edge correspondence then implies the existence of two counter-propagating edge states at the PhC's boundaries that cross the bandgap in a linear fashion.\:\cite{PhysRevLett.114.223901, Barik_2016, parappurathDirectObservationTopological2020} While any lattice defect in a PhC potentially hosts localized states, the promise of topologically protected edge states makes it especially interesting to study localized states of defects in PTIs.\:\cite{gaoDiracvortexTopologicalCavities2020, jalalimehrabadSemiconductorTopologicalPhotonic2020, sirokiTopologicalPhotonicsCrystals2017, sunTopologicalRingcavityLaser2021, yangTopologicalWhisperingGallery2018} As previous theoretical and experimental studies demonstrate, the helicity of the radiation emitted from both edge states is coupled to their unique (and opposing) pseudospins, allowing for spin-dependent excitation and probing of counterpropagating modes from the far field.\:\cite{PhysRevLett.114.223901, Barik_2016, andersonUnidirectionalEdgeStates2017, gorlachFarfieldProbingLeaky2018} In consequence, we can examine the optical and topological properties of the system using the Fourier-reflectometry setup schematically depicted in Figure \ref{fig:1}c,\:\cite{gorlachFarfieldProbingLeaky2018, parappurathDirectObservationTopological2020} whereby we employ two different beam paths to retrieve the cavity mode spectra and the individual modes' profiles in real- and momentum space (see the Experimental Section for a detailed description).
Figure \ref{fig:1}d shows a measurement of reflected intensity (obtained with the orange beam path in Figure \ref{fig:1}c) for a cavity that consists of a single unit-cell expanded defect within a shrunken bulk (see Figure \ref{fig:1}a). Without analyzer (Figure \ref{fig:1}d, left panel), the recorded signal has a bright background of directly reflected light and spectrally broad fringes (apparent at large $k_y$). The latter are a consequence of the sample geometry, which forms a Fabry-Pérot-like cavity between the suspended membrane and the underlying substrate (see Figure \ref{fig:1}a, inset).\:\cite{parappurathDirectObservationTopological2020} On top of this background, sharper resonant features appear, exhibiting Fano-type line-shapes along $k_y$ due to interference with the background radiation. These features include the dispersive bulk bands of the PhC surrounding the bandgap,\:\cite{gorlachFarfieldProbingLeaky2018, parappurathDirectObservationTopological2020} as well as flat bands at specific frequencies within the gap. We associate the latter with the cavity modes of the system in Figure \ref{fig:1}a, demonstrating that a single expanded unit cell supports a pair of confined modes around $208\,$THz with $Q\approx 600$.
To obtain the measurements in the center and right panel of Figure \ref{fig:1}d, we illuminate the sample with light linearly polarized along $X$ ($Y$) and detect the reflected signal with an orthogonal linear polarizer in the beam path, denoted by $X|Y$ ($Y|X$). This cross-polarization scheme allows us to better resolve the individual modes by suppressing background radiation, such that the resonances assume a typical Lorentzian lineshape. It relies on the cavity modes converting polarization of the incident to outgoing light fields, which as we will show is always possible for well-chosen alignment of the focused incident beam with respect to the cavity center.

\begin{figure*}[hbt]
  \centering
  \includegraphics[width=.75\linewidth]{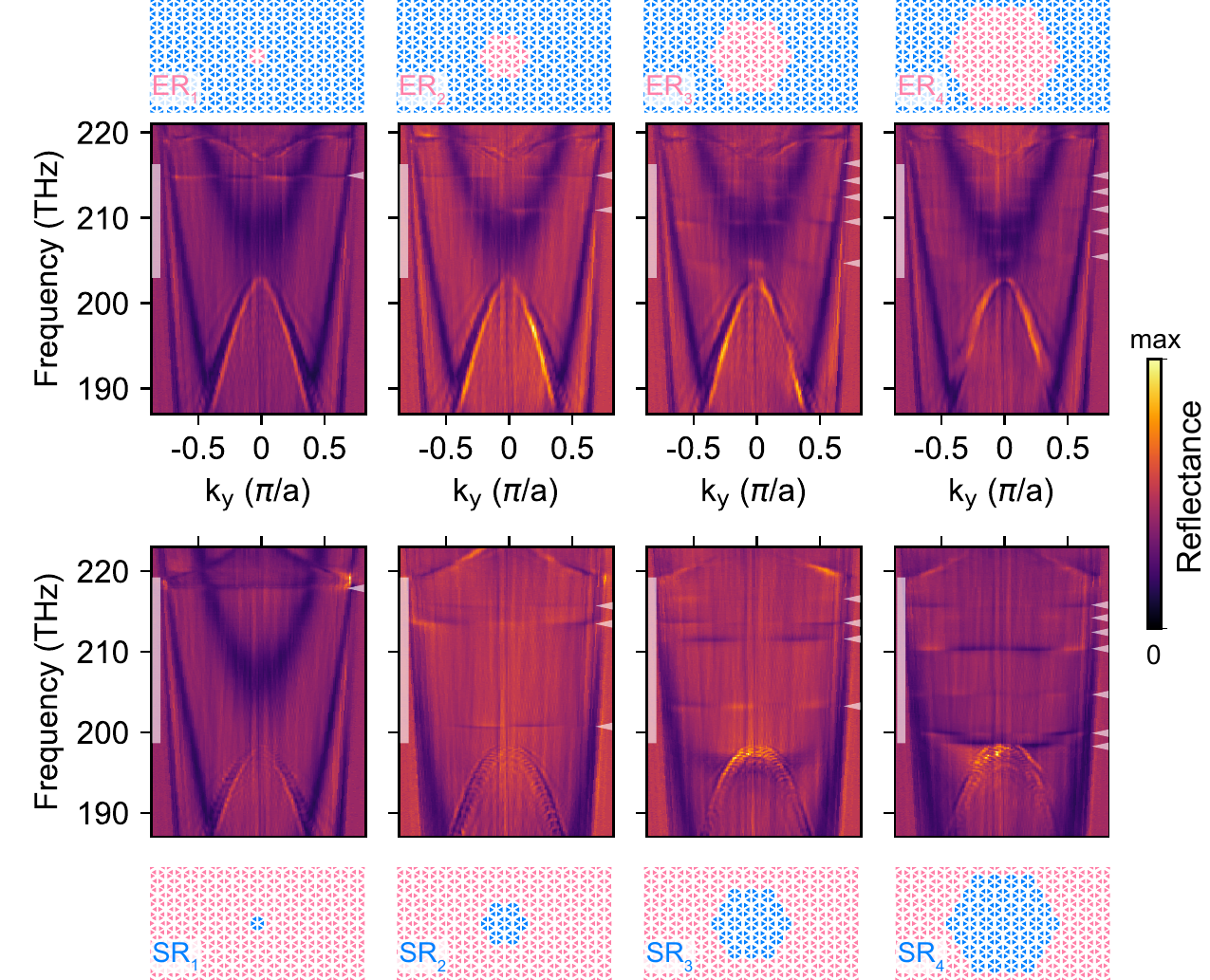}
  \caption{Dispersion vs. cavity size. Experimentally retrieved dispersion diagrams of hexagonal cavities formed by patches of expanded-in-shrunken (ER$_n$, upper row) or shrunken-in-expanded (SR$_n$, lower row) unit cells, where $n$ denotes the cavity's side length and the input is $Y$-polarized (no analyzer). Recognizable cavity modes as well as the bulk band gaps are highlighted by white pointers and light shaded bars, respectively, and the cavity geometries are depicted schematically.}
  \label{fig:2}
\end{figure*}

\subsection{Radiation profiles of topological cavity modes}

We employ reflectometry measurements to study the mode spectra for cavities of varying size and type. \figref{fig:2} displays the experimentally measured reflection of $X$-polarized light for expanded and shrunken defect cavities labeled ER$_n$ and SR$_n$, respectively, where $n$ denotes the hexagon's side length in terms of unit cells ($n=1\dots 4$). 
The number of states within the bandgap increases with $n$, whereby the expanded (Figure \ref{fig:2}, top row) and shrunken (Figure \ref{fig:2}, bottom row) resonators follow the same trend of decreasing free spectral range (FSR), exhibiting a comparable number of modes for a given cavity size. A reduction of FSR with cavity size is qualitatively expected as it generally scales inversely proportional to the mode volume. It must be noted that the modes we excite strongly depend on the vectorial field overlap with the focused input beam, hence, only a single mode is visible for ER$_1$ in Figure \ref{fig:2} (as opposed to Figure \ref{fig:1}d, left panel). Looking at the bands' intensity variations for different $k_y$, the resonant states at discrete frequencies display a distribution of emission angles characteristic of their multipolar nature. The latter is especially relevant in the context of topological lasing and generation of vortex beams, since it determines the symmetries and the amount of orbital angular momentum carried by the radiated fields.\:\cite{sunTopologicalRingcavityLaser2021} As a consequence of reciprocity, the multipolar order also determines the symmetries of the incident fields that can efficiently couple to the modes.

\begin{figure*}[hbt]
  \centering
  \includegraphics[width=.75\linewidth]{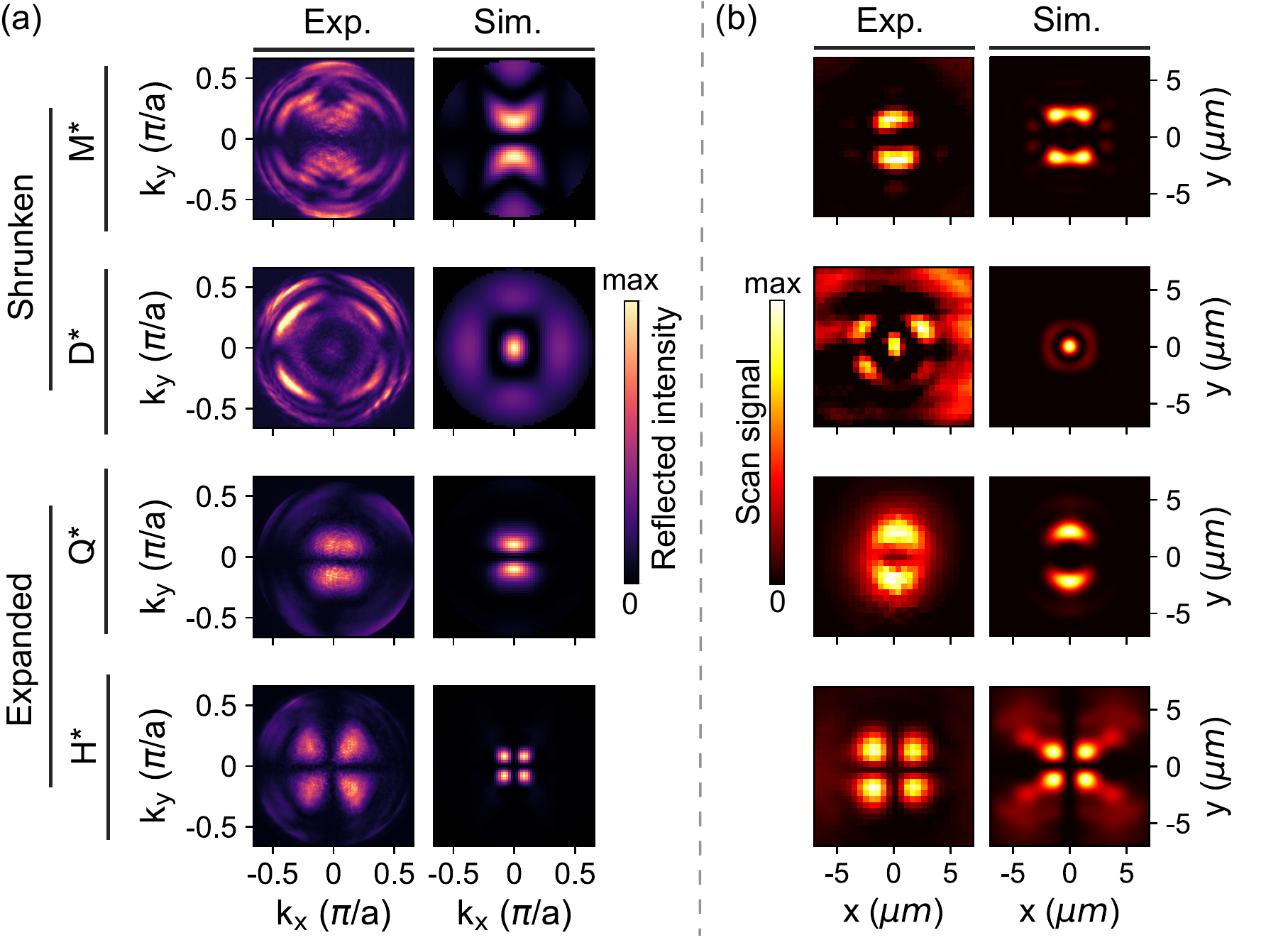}
  \caption{Angular and spatial cavity mode profiles.  a) Comparison of experimentally retrieved and numerically calculated radiation patterns for the M* (monopole), D* (dipole), Q* (quadrupole), and H* (hexapole) modes indicated in Figure \ref{fig:4}ab. In the experimental images, the polarizer and analyzer are aligned along $X$ and $Y$, respectively, and the simulated $X$-polarized field is shown. b) Spatial cavity mode profiles with polarizer and analyzer aligned along $Y$ and $X$, respectively, and the simulated $Y$-polarized field is shown.}
  \label{fig:3}
\end{figure*}

We therefore study the radiation patterns of individual modes by selectively addressing the resonances with a monochromatic light source (Figure \ref{fig:1}c, red beam path). We tune the wavelength of the incident light on resonance with the cavity mode of interest and record the reflected and re-radiated intensity in the back focal plane (BFP) of the microscope objective used for focusing and collection. In \figref{fig:3}a, we compare the experimental ($X|Y$) BFP patterns for selected modes in ER$_2$ and SR$_2$ cavities to numerical calculations. The latter are determined by analytically propagating the near fields obtained from finite-element-method calculations to the far field (see the Experimental Section for details on the numerical model). The orthogonal polarization projection as well as additional modes are shown in \figref{fig:FigS1} of the Supporting Information (SI). It must be noted that, due to the experimental focusing scheme, the cross-polarized configuration cannot suppress all incident light, which is present as a 4-fold symmetric background in the BFP.\:\cite{hongBackgroundFreeDetectionSingle2011} On top of this background, we see an excellent correspondence between calculated and recorded fields, sharing all characteristic symmetries (e.g. nodal lines). The number of radial nodal lines, together with the phase information in \figref{fig:FigS2} of the SI, serves as an indicator for the topological charge of the emitted vector beam, implying the possibility to generate light with a certain amount of orbital angular momentum. Experimental deviations in the intensity distributions may be attributed to the structural imperfections in the fabricated device as well as reflected stray light. Certain modes display considerably directional emission into a narrow spectrum of wavevectors around the normal direction (see also Figure \ref{fig:FigS2} of the SI) as a consequence of the topological edge state transport that involves Bloch components with well-defined, near-zero in-plane wavevectors, in contrast to trivial WGM resonators that rely on total internal reflection.\\ 
The exact modes we excite strongly depend on the position and polarization of the focused input beam, which is of particular relevance for applications that rely on the addressing of individual resonances. Therefore, we record polarization-resolved spatial cavity mode profiles and compare them to numerical predictions in Figure \ref{fig:3}b. In experiment, we raster-scan the input beam over a grid of positions whilst recording BFP images, similar to the ones in Figure \ref{fig:3}a. Each pixel in the reconstructed map then corresponds to the radiated intensity at that position, integrated over the BFP. In these maps, the cavity is centered around the origin. The numerical results are obtained by propagating the simulated near field to the far field, restricting the angular spectrum to the experimentally accessible range, and then performing the back-transformation. We present the results for excitation with $X$-polarized light (and cross-polarized detection) which, as can be inferred from reciprocity, correspond to the $X$-polarized radiation patterns in Figure \ref{fig:3}a and hence display the same symmetries (for a more detailed discussion and additional maps of other modes and polarization projections see Section \ref{sec:S1} and Figure \ref{fig:FigS2} of the SI). Measurements and numerical predictions correspond well to each other, successfully recovering all characteristic symmetries. It is worth noting that, apart from the dipolar modes, the radiated fields and spatial mode profiles possess a node at the origin. This implies that an angled, off-center excitation is essential to efficiently couple light into the system from the far field. We achieve this by focusing a Gaussian beam to obtain a broad spectrum of incident wavevectors which we translate across the sample plane. More efficient in-coupling could be achieved by means of structured illumination with input beams that are precisely tailored toward the cavity mode field of interest.\:\cite{rubinsztein-dunlopRoadmapStructuredLight2016} The good agreement between experiment and calculations allows for a straightforward categorization of the modes in terms of their multipolar nature, which we infer from the simulated near field phase (see Figure \ref{fig:FigS2} of the SI). We identify the modes in Figure \ref{fig:3} as monopolar (M), dipolar (D), quadrupolar (Q) and hexapolar (H). To elucidate their origin and scaling behavior, we continue by identifying all modes observed in the ER$_n$ and SR$_n$ cavities (up to $n=3$).

\begin{figure*}[hbt]
  \centering
  \includegraphics[width=\linewidth]{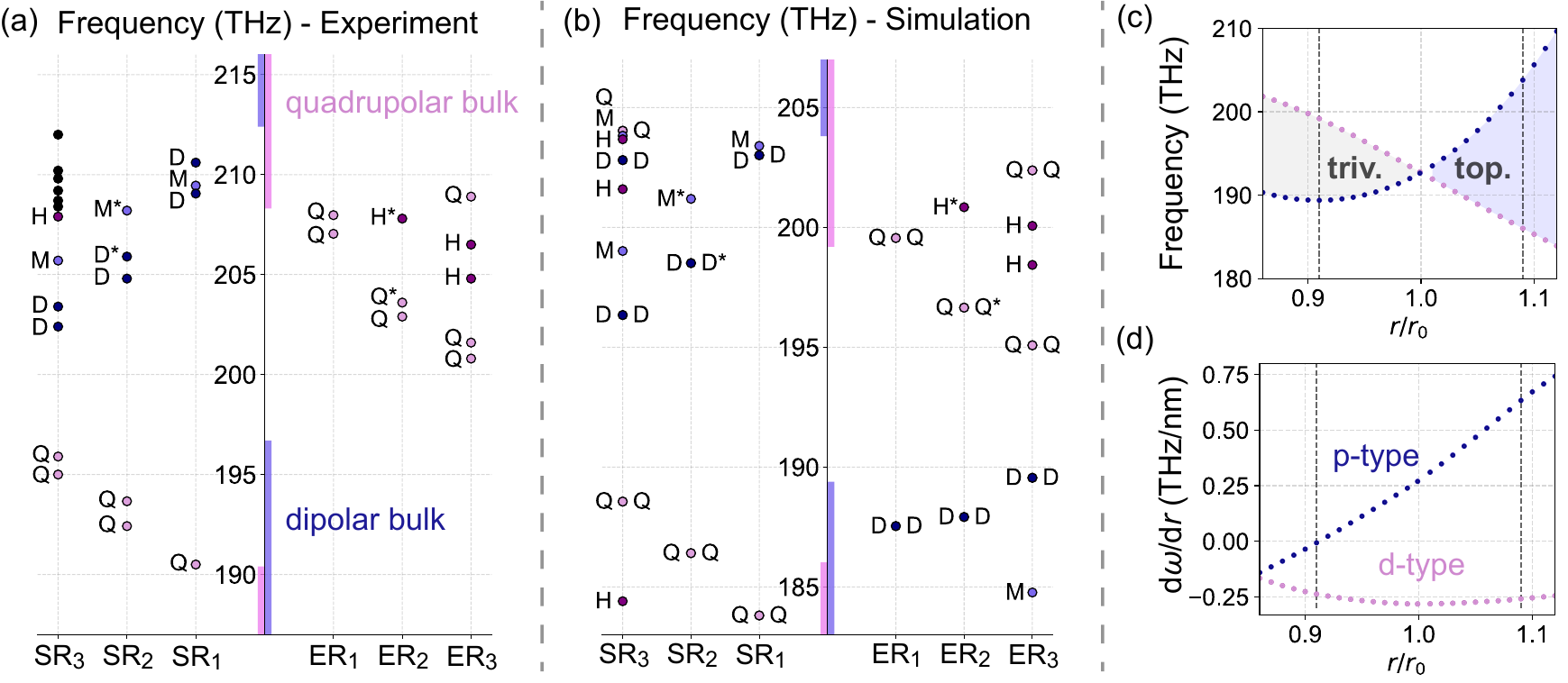}
    \caption{Nature of cavity modes.  a) Experimentally and b) numerically retrieved mode spectra for shrunken and expanded cavities of different sizes. Cavity modes are labeled according to their multipolar order, i.e. monopolar (M), dipolar (D), quadrupolar (Q) and hexapolar (H). Mode colors indicate the bulk bands from which they detach: quadrupolar for Q and H modes (pink shades), dipolar for M and D modes (purple shades). Vertical bars indicate the dipolar (p-type, purple) and quadrupolar (d-type, pink) bands of the bulk lattice surrounding the cavity (expanded for SR$_n$, shrunken for ER$_n$). c) Numerically retrieved frequency of the top and bottom bulk bands as a function of the unit cell expansion factor ($r/r_0$). Band inversion takes place upon transformation from shrunken (trivial) to expanded (topological) geometry. The distance of the triangular air holes' centroids from the unit cell center is denoted by $r$, with ${r=r_0}$ for an ordinary lattice (see Figure \ref{fig:1}b). Our choice of the unit cell expansion factor for the shrunken ($0.91$) and expanded ($1.09$) case are indicated by dashed lines in c) and d). d) Frequency change for an infinitesimal expansion, as obtained from first-order perturbation theory.}
  \label{fig:4}
\end{figure*}

\subsection{Scaling behavior of mode spectra}

The experimentally extracted center mode frequencies and multipolar orders are displayed in \figref{fig:4}a. We highlight the bandgaps in the region between the p- and d-like bulk bands. For the shrunken and expanded case, the gaps have relative sizes of $5.7\,\%$ and $10.9\,\%$, centered around $202.5\,$THz and $201.4\,$THz, respectively. This is quantitatively well reproduced by numerical calculations. We notice a constant offset on the order of $\sim 5-8\,$THz that is attributed to deviations of the real device from the ideal design. 
The spatial symmetries of the PhC determine the symmetries of the modes it potentially hosts. Our cavities can be regarded as defects in an otherwise unperturbed TE-like PhC slab, where the surrounding lattice as well as the defect obey a $C_6$ rotational symmetry. Solving the electromagnetic eigenvalue equation governing the system shows that such a lattice supports precisely the types of modes shown in Figure \ref{fig:3} (M, D, Q and H), whereby the spherical multipoles are named according to the respective rotational symmetries of the $H_z$ field.\:\cite{joannopoulosPhotonicCrystalsMolding2008}\\
Rotations of $60^{\circ}$ transform eigenstates into each other and may be used to construct two mutually orthogonal bases for $D$ and $Q$. This implies double degeneracy and is in contrast to $M$ and $H$, which are singlet states.\:\cite{se-heon_kim_symmetry_2003} The degenerate $D$ and $Q$ modes can be interpreted as traveling-wave WGMs, i.e., a superposition of two counter-propagating edge states traveling along the resonator's perimeter with equal amplitude but different phase. The $M$ and $H$ modes, however, have no traveling-wave analogue, and the system is better described as a PhC defect cavity that supports collective resonances extending throughout its bulk. These standing wave modes (SWMs) appear due to the reduced spatial symmetry of the resonators.\:\cite{luSelectiveEngineeringCavity2014} While conventional ring cavities may be designed perfectly circular to only feature WGMs and no SWMs, cavities in the PTI platform, regardless of their size, inevitably break continuous rotational symmetry and display both types of modes. To control and reduce the appearance of SWMs, we match the symmetries of the cavities to the $C_6$ symmetry of the unit cell by a hexagonal design.\:\cite{yuCriticalCouplingsTopologicalinsulator2021}
The consequences of these spatial symmetry considerations for the mode spectra are nicely corroborated by the ideal numerical model presented in Figure \ref{fig:4}b, displaying doubly degenerate $Q$ and $D$ WGMs as well as non-degenerate $M$ and $H$ SWMs. In contrast to simulations, we observe splitting on the order of $\lesssim  1\,$THz for $D$ and $Q$ modes in the fabricated device. It is known that a small splitting in straight QSHE-based topological PhC edge state waveguides exists as a consequence of $C_6$ symmetry being only approximately fulfilled at a straight interface between shrunken and expanded unit cells, causing the two pseudospin states to mix and a small spin-spin-scattering gap to open up.\:\cite{PhysRevLett.114.223901} However, since a hexagonal cavity obeys $C_6$ symmetry, a strict degeneracy between modes of different polarization is expected in theory.\:\cite{se-heon_kim_symmetry_2003} Notably, numerical simulations of larger hexagonal cavities have not always confirmed this degeneracy,\:\cite{sirokiTopologicalPhotonicsCrystals2017} possibly due to simulation imperfections. In finite-element-method calculations where we ensure $C_6$ symmetry is obeyed in all aspects of the simulation, we indeed observe that splitting is negligibly small (see Figure \ref{fig:4}b).\\
If we compare the spectra for increasingly large expanded or shrunken resonators, we observe that, while additional modes emerge in the bandgap, their frequencies shift in qualitatively and quantitatively different ways. It furthermore becomes evident, especially for the simulation results in Figure \ref{fig:4}b, that these modes originate and detach from the bulk bands upon increasing the cavity size. Strikingly, if we consider their multipolar order, we observe an inverted scaling behavior of the modes for the shrunken and expanded case. For SR cavities, $M$ and $D$ modes appear to originate from the top bulk bands and decrease in energy with cavity size, while $Q$ and $H$ modes emerge from the bottom bulk and shift upwards. The situation is reversed for the expanded case, where instead $Q$ and $H$ modes shift downwards and $M$ and $D$ upwards, as evidenced by simulations. We attribute the lack of $M$ and $D$ modes for ER cavities in experiments to their frequencies falling into the lower bulk region, causing hybridization with extended PhC slab modes and thus increasing in-plane radiation losses, making these resonances harder to resolve in measurements. Apart from ordinary photonic bandgap confinement, band inversion induced confinement in the topological PhC contributes to the localization of cavity modes in the bulk band region (e.g. $M$ in ER$_3$, or $Q$ in SR$_1$).\:\cite{shaoHighperformanceTopologicalBulk2020b}\\
The scaling of mode frequencies becomes more clear if we trace the upper and lower band edges for a continuous transformation from shrunken to expanded geometry, as shown in Figure \ref{fig:4}c. We vary the unit cell expansion factor $r/r_0$ and thus track the process of band inversion upon which p- and d-type bands switch their role and become degenerate for the perfectly triangular lattice with $r=r_0$. We indicate the radii of our realizations of shrunken and expanded geometry and denote the trivial or topological nature of the bandgap, which the p-type bands cross linearly while the d-type modes feature a rather parabolic progression. The curves for the degenerate pairs lie exactly on top of each other. This behavior implies that, if we start out with a lattice of a specific type and introduce defects in the shape of unit cells of the other type, we effectively dope it with a material with inverted bandstructure. If we, for instance, consider an infinite shrunken lattice ($r<r_0$) into which we introduce a point defect in the shape of an expanded cell (ER$_1$), this defect ``pulls'' and ``pushes'' modes from the upper and lower bulk bands, respectively. In the limit where we replace every shrunken unit cell by an expanded one, this process ultimately inverts the bandstructure. For finite patches of increasing size however, as in our case, we observe discrete modes (with decreasing FSR) whose frequencies evolve in opposite directions for ER and SR systems.\\
To support this argument, we employ first order perturbation theory\:\cite{joannopoulosPhotonicCrystalsMolding2008, johnsonPerturbationTheoryMaxwell2002} and determine the frequency shift $\mathrm{d}\omega$ upon infinitesimal variation of the radius $\mathrm{d}r$ through suitable integration of the mode fields (for details see Section \ref{sec:S2} and \figref{fig:FigS3} of the SI). 
The result is shown in Figure \ref{fig:4}d and, as expected, represents the derivative of the graphs in Figure \ref{fig:4}c with respect to $r$. While the p-type modes display a fairly constant rate of energy change, corresponding to the linear crossing of the bandgap in Figure \ref{fig:4}c, the parabolic progression of the d-type modes therein manifests as linear slope of the respective curves in Figure \ref{fig:4}d (the degenerate mode pairs lie on top of each other again). The frequency scaling of the bands underlines the importance of choosing sufficiently different expansion factors for the defect and bulk unit cells in order to achieve comparable bandgap sizes and sufficient mode confinement, especially visible for the frequencies of the d-type bands which vary only weakly around $r/r_0=0.91$ (shrunken lattice). The sign of $\mathrm{d}\omega/\mathrm{d}r$ is reflected in the direction of frequency change for the various multipoles in Figure \ref{fig:4}ab, further underlining the qualitatively different behavior in expanded and shrunken systems. The inversion of mode scaling behavior results from the band inversion underlying the topological phase transition in the platform and represents a design advantage over trivial PhC lattice defect or WGM resonators, as it offers an additional degree of freedom to strongly tailor cavity mode spectra by small variations to the unit cell geometry while preserving the overall footprint.

\begin{figure*}[ht]
  \centering
  \includegraphics[width=\linewidth]{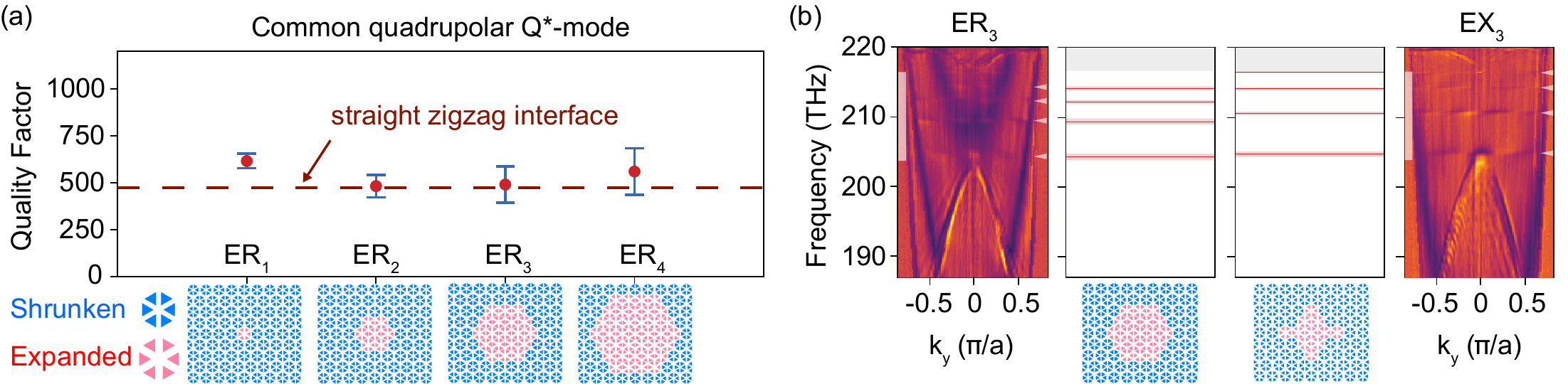}
  \caption{Fingerprints of non-trivial topology. a) Experimentally extracted quality factors for the Q* mode common to all investigated ER cavities, averaged from seven line cuts to the angularly resolved mode spectra and given alongside with the standard error. The red dashed line indicates the radiative quality factor for a straight zigzag-type topological interface. b) Comparison of modes for cavities of similar perimeter but strongly differing shape, as indicated by pictograms. In measurements, the cavity modes and bulk band gaps are highlighted by white pointers and light shaded bars, respectively. Extracted center mode frequencies and corresponding linewidths are presented in the central panels as shaded horizontal lines.}
  \label{fig:5}
\end{figure*}

\subsection{Size- and shape-dependence of leakage radiation}

Another key advantage of topological platforms with respect to conventional systems is considered to be their enhanced robustness against defects. In \figref{fig:5}, we investigate the influence of cavity size and shape on the mode spectra to search for hallmarks of topological protection. 
An obvious point to study is the optical loss, which we compare for cavities of increasing size. To this end, we choose to examine the higher energy quadrupolar mode (Q*) which all expanded cavities (ER$_{1-4}$) have in common (see Figure \ref{fig:2}, \ref{fig:3}, \ref{fig:4}ab). For each resonator, we fit a model of a general resonance lineshape (see \eqrefrb{eq:1} in the Experimental Section) to the normalized reflection in order to extract the mode frequency and linewidth from seven cross cuts along $k_y$.  The average extracted quality factors alongside the standard error of the mean are shown in Figure \ref{fig:5}a. We observe no significant variation of the quality factor with size, suggesting that scattering due to the curvature of the perimeters is not the dominant loss mechanism for these cavities, even for the smallest ones. Since the experimental linewidths are comparable to those of regular edge state dispersion curves along a straight zigzag-type interface (Figure \ref{fig:5}a, red dotted line),\:\cite{parappurathDirectObservationTopological2020} the states' intrinsic leakage radiation seems largely unaffected by the curvature of the interface and is likely to be the dominant loss-mechanism for our resonators. 
If we, for comparison, consider the same situation in topologically trivial WGM cavities,\:\cite{joannopoulosPhotonicCrystalsMolding2008} there is in general a trade-off between mode volume and optical losses. In small cavities (i.e., cavities with small mode volumes), light is forced onto a strongly curved path; a ring-shaped waveguide that is dominated by bending losses. The latter are expected to disappear in the limit of infinitely large cavities, as the perimeter effectively approaches a straight waveguide. In absence of any other loss mechanisms the quality factor continuously increases with mode volume. However, in practice it is limited by processes such as scattering from fabrication imperfections and intrinsic material losses.  Therefore, we expect the quality factor to increase with size until eventually converging to the value for a straight waveguide.
In the QSHE-type PhC cavities, however, the fairly constant quality factor implies that non-trivial topology protects the edge states from scattering at the cavities' corners, to within the precision given by the intrinsic radiation quality factor.\\
The robustness of cavity modes can also be studied by deliberately introducing more corners whilst keeping the size constant. Figure \ref{fig:5}b shows a comparison between the dispersion of the hexagonal ER$_3$ cavity and  another cavity EX$_3$, with same perimeter but drastically different shape. If we compare the extracted mode frequencies and linewidths, we notice that, apart from an overall increase in FSR, the mode spectra look fairly similar and the number of modes remains the same. Furthermore, while the quality factors for ER$_3$ ($Q\gtrsim 210$) are actually slightly lower than for the perturbed cavity EX$_3$ ($Q\gtrsim 260$), they remain on the same order of magnitude. The number of modes as well as their lifetime is considerably well retained even for a significant perturbation of the resonator's shape, which can be regarded as a further hallmark of topological protection. Moreover, the leakage radiation rate is a fundamental property that is not significantly affected by cavity size and shape.

\section{Conclusion}

In conclusion, we experimentally investigate the optical confinement and radiation of telecom light in leaky topological PhC cavities mimicking the QSHE via Fourier spectropolarimetry and corroborate our measurements with full-wave finite-element-method calculations. We recognize the presence of both traveling WGMs and extended PhC defect SWMs, and identify them as the ideally degenerate $D$/$Q$ and non-degenerate $M$/$H$ multipoles. We elucidate the origin and scaling behavior of the multipolar modes supported by the system and relate it to the angularly resolved radiation patterns and spatial cavity mode profiles, providing a route toward selectively addressing individual modes by spectral, spatial, angular, and polarimetric means. Finally, we discuss the implications of the simultaneous presence of radiative loss and non-trivial topology onto the modes' properties, demonstrating the robustness of cavity spectra and quality factors with respect to alterations of shape and size, in sharp contrast to conventional, topologically trivial micro- and nanocavities. The key advantage of robustness alongside other inherent properties like radiation in vortex beam modes make them promising building blocks for applications in integrated passive and active nanophotonic devices.\:\cite{iwamotoRecentProgressTopological2021}  Although our study focuses on the technologically relevant telecommunication band, these findings may be applied more broadly in the context of lossy topological bosonic systems, including other frequency regimes as well as mechanical and phononic platforms. As such, we envision the results of our study to advance the development of novel functional devices that aim to manipulate and control classical and quantum information.

\section{Experimental Section}

\paragraph{Simulations:}
Full-wave finite-element-method simulations in 3D were performed using the COMSOL Multiphysics RF Module.\:\cite{COMSOL52a} The refractive index of silicon was set to $n = 3.48$, with a slab thickness of $220\,$nm. The unit cell consisted of equilateral triangular air holes, with a triangle side length of $s = 250\,$nm and a lattice constant of $a = 800\,$nm. By adding perfectly matched layers above the simulation box, we retrieve the linewidth of the (quasinormal) eigenmodes, defined as two times the imaginary part of the complex eigenfrequency. The simulation domain was chosen to be hexagonal to match the $C_6$ symmetry of the system, and terminated by scattering boundary conditions. We extract the near field profiles on a regular grid in a plane located $20\,$nm above the slab. The near field phase enables us to classify the cavity modes according to their multipolar nature, and the full field information is used to retrieve the far field radiation patterns and spatial cavity mode profiles, as detailed in the main text. \medskip

\paragraph{Device fabrication:} 
The PhC slab was fabricated on a silicon-on-insulator platform with a $220\,$nm thick silicon layer on a $3\,\mathrm{\mu m}$ buried oxide layer. The fabrication was performed in two steps: First, a positive electron-beam resist of thickness $240\,$nm (AR-P 6200.09) was spin-coated between a monolayer of adhesion reagent HMDS and a conductive layer of E-Spacer 300Z. Then, the PhC design was patterned in the resist using e-beam lithography on a Raith Voyager with $50 \,$kV beam exposure. The e-beam resist was developed in pentyl-acetate/o-Xylene/MIBK:IPA(9:1)/isopropanol, and the chip subsequently underwent reactive-ion etching in HBr and O$_2$. Finally, the buried-oxide layer was removed in an aqueous 5:1 solution of hydrofluoric acid for $19\,$min and the sample was then subjected to critical point drying in order to obtain free-standing PhC membranes.\:\cite{Barik_2016} The PhC lattice design features a honeycomb configuration of equilateral triangles (side length $s=0.3125\cdot a$) in a hexagonal unit cell with lattice constant $a$. The unit cell expansion factor $r/r_0$ was chosen to be $0.91$ and $1.09$ for the shrunken and expanded case, respectively, whereby $r$ is the distance of the centroids of the triangular air holes from the unit cell center, with $r=r_0$ for a perfectly triangular lattice. Two samples were fabricated, with $a = 800\,$nm in Figure \ref{fig:2} and \ref{fig:5}, and $a = 852\,$nm in all other experimental images. \medskip

\paragraph{Experimental setup:} 
To measure the photonic dispersion, we use a $200\,$mW supercontinuum source (Fianium WhiteLase Micro) that generates light with a broadband spectrum. Its output is filtered by a long-pass filter with a cutoff wavelength of $1150\,$nm and coupled into a single-mode optical fiber. The IR light from the fiber is collimated by an achromatic lens and passed through a linear polarizer (LP) and an achromatic quarter-wave plate (QWP), which together define the polarization of the input beam (PO1). A beam splitter (BS) steers the input light to an aspheric objective (Olympus LCPLN50XIR, $50\times$, numerical aperture $= 0.65$), which focuses the incident Gaussian beam onto the sample. In order to precisely position the sample in the focal plane, it is attached to a XYZ-movable piezo actuator (MCL Nano-3D200FT, controlled via MCL ND3-USB163), which itself is mounted atop a manual XYZ translation stage for coarse alignment. Reflected light is collected by the same objective and passed through the BS and a second set of LP and QWP (PO2) to project the BFP radiation onto the desired polarization state. It then passes a Fourier lens (FL) which, together with a tube lens (TL), images the objective's BFP onto the entrance slit of a spectrometer (Acton SpectraPro SP-2300i). Optional custom spatial filters (SFs) are placed in the image plane between the FL and TL to define the sample area from which light is collected and to suppress stray light. The (vertical) entrance slit of the spectrometer is aligned with the optical axis and selects a cross-cut along $k_x = 0$ in the reciprocal plane, confirmed using a test grating sample. With the help of two parabolic mirrors (PMs) for focusing and collection, the spectrometer grating then disperses the broadband IR light orthogonally to the slit, such that the InGaAs IR camera (AVT Goldeye G-008 SWIR) placed at the spectrometer output records images of frequency versus $k_y/k_0$, where $k_0$ is the free-space wave vector. The wave vector resolution is $\delta k_y/k_0 \approx 0.009$, and the typical spectral resolution is $\sim 87\,$GHz. For recording the radiation patterns, we use a monochromatic laser source (Toptica Photonics CTL 1500) and remove a flip mirror (FM) to guide the reflected signal through another TL and directly image the full angularly resolved BFP onto a second (similar) InGaAs IR camera, instead of imaging it onto the spectrometer slit. \medskip

\paragraph{Extraction of cavity mode frequencies and quality factors:} 
In order to extract the center frequencies and quality factors of the cavity modes from cross-cuts of the normalized reflection, we fit a set of general (Fano) resonance lineshapes of the form
\begin{equation}
\label{eq:1}
R(\omega) = \left| A_0 + \sum^n_{j=1} A_j e^{i\phi_j} \frac{\gamma_j}{\omega - \omega_{0_j} + i\gamma_j} \right|^2 
\end{equation}
where $A_0$ is a constant background amplitude; and $A$, $\phi$, and $\omega_0-i\gamma$ are the amplitude, phase, and complex frequency of individual Lorentzians, respectively. For cross-polarized measurements $n=1$, i.e., the lineshape is well reproduced by a single Lorentzian due to negligible $A_0$. For reflection measurements recorded without analyzer $n=2$, whereby one of these Lorentzians models the cavity mode, while the second (broad) Lorentzian accounts for the slowly varying background reflection. Quality factors are defined as $Q_j = \omega_0 / (2\gamma_j)$. \vspace{\baselineskip}

\textbf{Acknowledgements} \par 
R.B. and N.P. fabricated the devices and carried out the far field measurements. R.B. and N.P. performed data analysis and modeling. R.B. drafted the manuscript. E.V. and L.K. conceived and supervised the project. All authors contributed extensively to the interpretation of the results and the writing of the manuscript.
This work is part of the research program of the Netherlands Organisation for Scientific Research (NWO). The authors acknowledge support from the European Research Council (ERC) Advanced Investigator Grant no. 340438-CONSTANS and ERC starting grant no. 759644-TOPP.

\clearpage


\setcounter{section}{0}
\setcounter{equation}{0}
\setcounter{figure}{0}
\setcounter{table}{0}
\setcounter{page}{1}
\renewcommand{\theequation}{S\arabic{equation}}
\renewcommand{\thefigure}{S\arabic{figure}}
\renewcommand{\bibnumfmt}[1]{[S#1]}
\renewcommand{\citenumfont}[1]{S#1}
\renewcommand{\theHequation}{Supplement.\theequation}
\renewcommand{\theHtable}{Supplement.\thetable}
\renewcommand{\theHfigure}{Supplement.\thefigure}

\makeatletter
\let\@keywords\@empty
\makeatother

\title{Supplemental material:\\ Interplay of leakage radiation and protection in topological photonic crystal cavities}

\maketitle
\pagebreak
\onecolumngrid

\section{Radiation patterns and spatial cavity mode profiles}
\label{sec:S1}

\begin{figure}[ht]
\centering
\includegraphics[width=\textwidth]{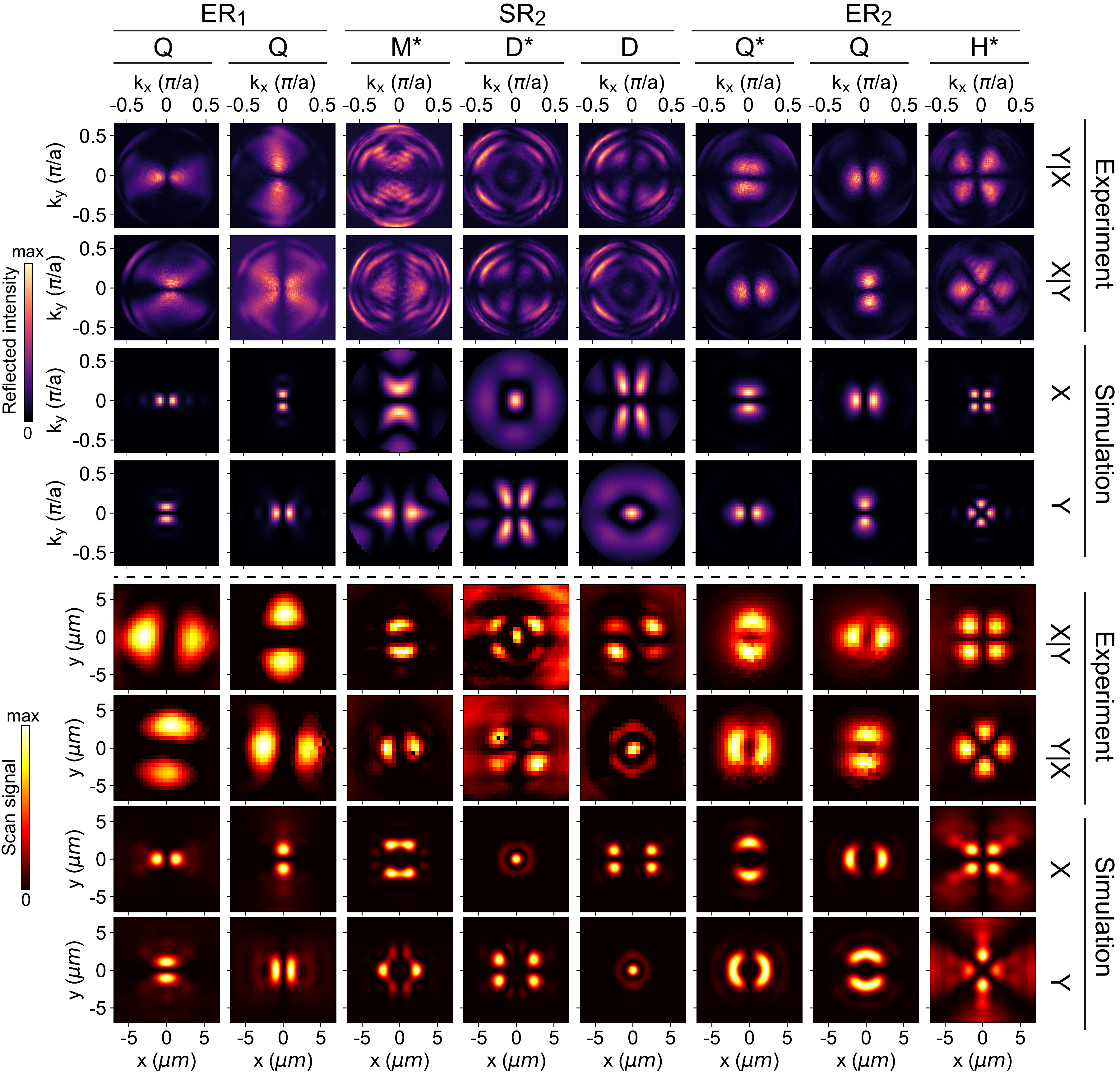}
\caption{Radiation patterns and spatial cavity mode profiles for various cavity modes in ER$_1$, SR$_1$, ER$_2$, and SR$_2$ (see Figure \ref{fig:3} and \ref{fig:4}ab of the main text).  The settings of polarizer and analyzer in experiments and the linear polarization of the numerical data are indicated, as well as the monopolar (M), dipolar (D), quadrupolar (Q), and hexapolar (H) nature of the modes. $M^*$, $D^*$, $Q^*$, and $H^*$ denote the same modes as in Figure \ref{fig:3} and \ref{fig:4}ab of the main text. Each subfigure is normalized to its respective maximum for reasons of visibility.}
\label{fig:FigS1}
\end{figure}

\begin{figure}[ht]
	\centering
	\includegraphics[width=\textwidth]{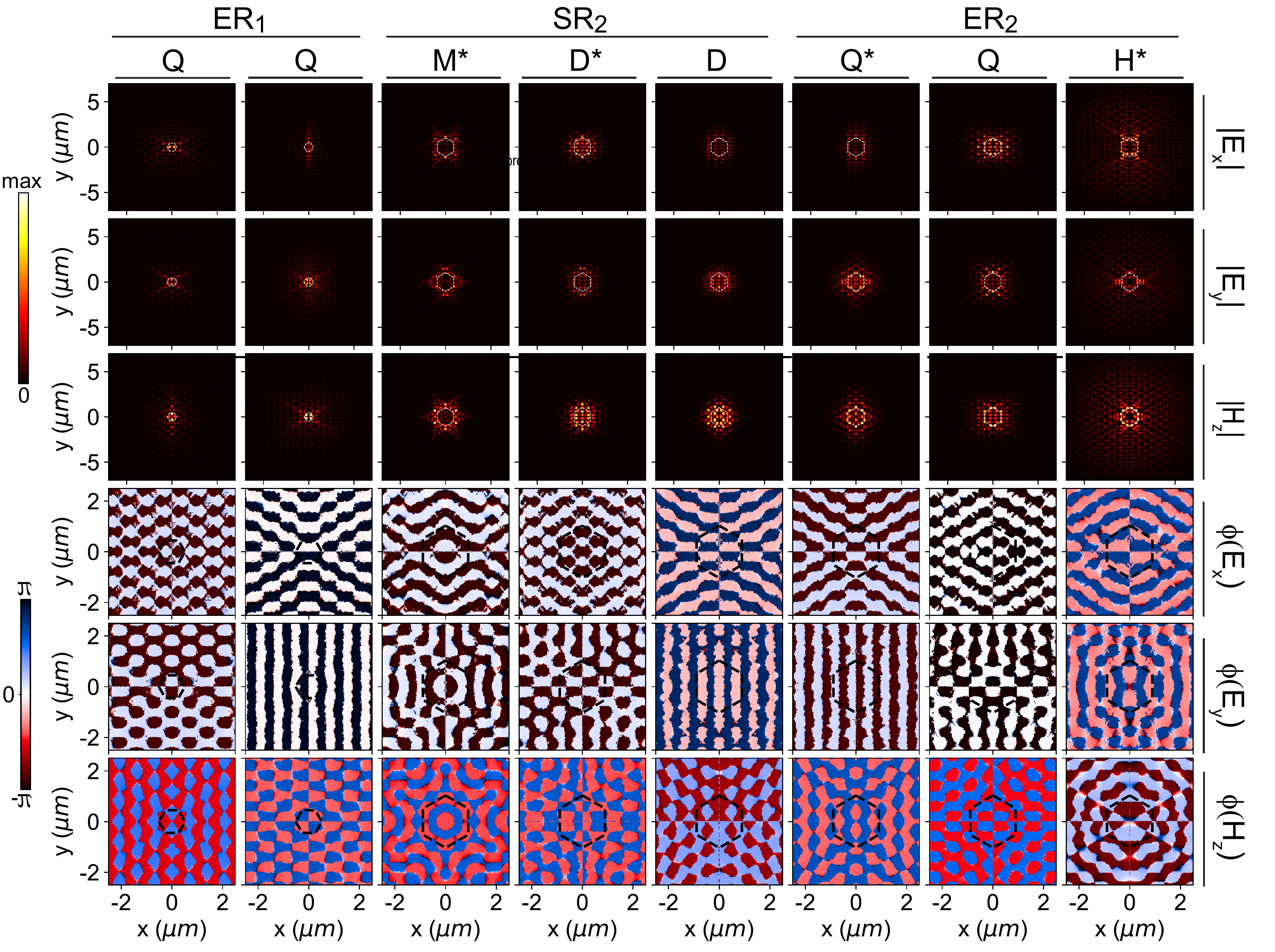}
	\caption{Simulated magnitude and phase of the non-zero field components in the symmetry plane of the TE type photonic crystal for the cavity modes shown in Figure \ref{fig:FigS1}. The hexagonal outlines of the cavities are indicated by dashed lines. Note that for reasons of visibility the top and bottom panel have different axes scalings.}
	\label{fig:FigS2}
\end{figure}

In the top panel of \figref{fig:FigS1}, we compare polarization-resolved experimentally and numerically retrieved radiation maps (see Figure \ref{fig:2} of the main text) for a selection of modes that is partly included in Figure \ref{fig:3} of the main text, encompassing every multipole present in our system. The measured back focal plane patterns correspond well to the simulated radiation profiles, retrieving all characteristic symmetries of the individual multipolar resonances. By comparing these to the numerically obtained near field maps of the non-zero field components in the symmetry plane of the TE type photonic crystal ($E_x$, $E_y$, $H_z$) shown in \figref{fig:FigS2}, we can infer the multipolar order of the probed cavity modes. Apart from the dipolar resonances ($D$/$D*$), both in-plane components of the modes' far fields feature radial nodal lines which, together with the phase information in Figure \ref{fig:FigS2}, shows that they represent vector beams that carry one (monopoles $M$ and quadrupoles $Q$) or two (hexapoles $H$) quanta of orbital angular momentum. As such, reciprocity implies most efficient excitation of these modes with similar vectorial input beams, such as higher order Laguerre-Gaussian beams. Reciprocity also dictates the relation between the symmetries of the angularly-resolved far field radiation and the spatial profiles of the cavity modes shown in the bottom panel of the figure.
The spatial mode profile is determined by calculating the excitation efficiency $\vec{\eta}$ of a cavity mode $\vec{E}_{cav}$ under illumination with the incident field $\vec{E}_{in}$ propagating along $z$. We notice that, if the wavefronts of both fields have the same frequency and the complex field profiles are well-matched in a certain $z$-plane, they remain well-matched during further propagation (i.e., $\eta$ is conserved during free-space propagation). 
We can thus, without loss of generality, consider the mode overlap in the symmetry-plane of the photonic crystal ($z=0$) and omit the $z$-dependence in the observables. For the TE-polarized guided modes in our photonic crystal it holds that $\vec{E}_{cav, z}=0$ in the symmetry-plane, i.e., the cavity mode is purely transverse for $z=0$. Furthermore, our radially symmetric focusing scheme does not admix the orthogonally polarized field components along $x$ and $y$. In our calculations of $\vec{\eta}$ we can therefore decompose the vectorial observables into their Cartesian in-plane components along $j=x,y$ and consider them separately. In experiments, we record the spatially resolved $\eta_j (x,y)$ by sweeping the position of the incident beam while assuming the cavity mode to remain centered around the origin, i.e.,

\begin{equation}
\eta_j (x,y) = \left| \iint \mathrm{d}x'\mathrm{d}y'\, E_{cav,j}^*(x',y')\, E_{in,j} (x'-x, y'-y) \right|^2 \,.
\end{equation}

Note that, for simplicity, we omit the commonly performed normalization of $\eta$ w.r.t.\ the total power. To gain further insight into the symmetry relations between the fields and the recorded signal let us first consider the complex quantity 

\begin{equation}
\eta'_j (x,y) = \iint \mathrm{d}x'\mathrm{d}y'\, E_{cav,j}^*(x',y')\, E_{in,j} (x'-x, y'-y) \,,
\end{equation}

such that $\eta_j = |\eta'_j|^2$. By applying a two-dimensional Fourier transform in consideration of the convolution theorem, taking the absolute squared of the resulting equation, and performing a back-transformation to the spatial domain, we finally arrive at an expression for the excitation efficiency

\begin{equation}
\eta_j (x, y) =  \mathcal{F}^{-1} \left\{ \left| \hat{E}_{cav,j}^*(k_x, k_y)\, \hat{E}_{in,j} (k_x, k_y) \right|^2 \right\} \,,
\end{equation}

where $\hat{E}_{cav,j}$, $\hat{E}_{in,j}$ are the spectral representations of the cavity mode and input field, respectively, and $\mathcal{F}^{-1}$ denotes the two-dimensional inverse Fourier-transform. The Fourier transform (and its inverse) commutes with orthogonal symmetry transformations such as rotations and reflections, and we can thus directly infer that $\eta_j$ has the same symmetries as the intensity of the product of $\hat{E}_{cav,j}$ and $\hat{E}_{in,j}$. Since in our case the input field is represented by a rotationally invariant Gaussian beam, $\eta_j$ retains the discrete symmetries of $\hat{E}_{cav,j}$. This fact is corroborated by the back focal plane patterns and spatial cavity mode profiles presented in Figure \ref{fig:FigS1}.

\section{Scaling behavior of mode spectra}
\label{sec:S2}

\begin{figure}[ht]
	\centering
	\includegraphics[width=0.5\textwidth]{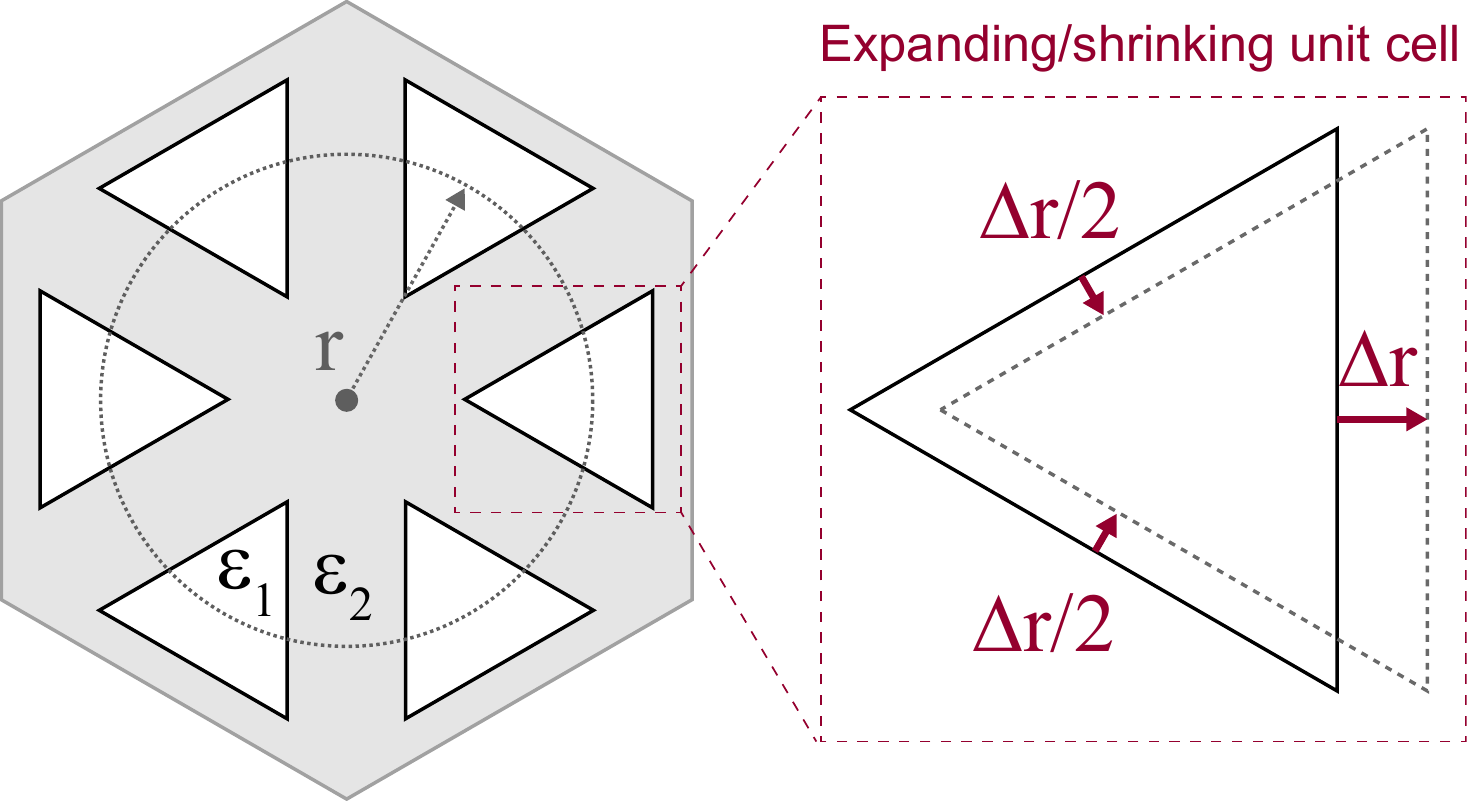}
	\caption{Sketch of an infinitesimal expansion/contraction of the unit cell, displaying the geometrical parameters used for a perturbative approach to calculate the scaling behavior of cavity mode frequencies with size, as shown in Figure \ref{fig:4}d of the main text.}
	\label{fig:FigS3}
\end{figure}

In order to elucidate the origin and scaling behavior of the modes for cavities of increasing size shown in Figure \ref{fig:4}ab of the main text, we treat an expanded (shrunken) cavity as a defect inside the surrounding shrunken (expanded) bulk lattice, constituted of perturbed unit cells with an increased (decreased) radius of the centroids of the triangular air holes $r$. We apply perturbation theory to study the effect of altering the unit cell expansion factor $r/r_0$, which may be viewed as a perturbation $\Delta\epsilon$ of the dielectric function. The frequency shift $\Delta\omega$ for an eigenmode of a system subject to such a perturbation may be written as \cite{joannopoulosPhotonicCrystalsMolding2008_SI}

\begin{equation}
\Delta\omega = -\frac{\omega}{2}\frac{\int\mathrm{d}^3\vec{r}\, \Delta\epsilon (\vec{r})\, |\vec{E} (\vec{r})|^2}{\int\mathrm{d}^3\vec{r}\,\epsilon (\vec{r})\, |\vec{E} (\vec{r})|^2} + \mathcal{O}(\Delta\epsilon^2) \,,
\end{equation}

where $E$ and $\omega$ denote the electric field distribution and frequency, respectively, of the unperturbed dielectric function $\epsilon$. Although this expression is applicable for a wide range of small perturbations, small displacements of dielectric boundaries with high refractive index contrast such as in our system represent moving discontinuities to which the formula cannot be applied. In this case, if we consider a dielectric boundary shifting from $\epsilon_1$ toward $\epsilon_2$ by a distance $\Delta r$ perpendicular to the interface (see \figref{fig:FigS3}), the correct expression is given by \cite{johnsonPerturbationTheoryMaxwell2002_SI}
\begin{equation}
\label{eq:dw_pert}
\Delta\omega = -\frac{\omega}{2}\frac{\int\mathrm{d}^2\vec{r}\, \left[ (\epsilon_1 - \epsilon_2)|\vec{E_{\parallel} (\vec{r})}|^2 - (\epsilon_1^{-1}- \epsilon_2^{-1}) |\epsilon \vec{E_{\perp}}(\vec{r})|^2 \right]}{\int\mathrm{d}^3\vec{r}\,\epsilon (\vec{r})\, |\vec{E} (\vec{r})|^2} \cdot \Delta r + \mathcal{O}(\Delta r^2) \ .
\end{equation}

Here, $\vec{E_{\parallel}}$ and $\vec{E_{\perp}}$ are the electric field components parallel and orthogonal to the boundary, respectively. We use \eqrefrb{eq:dw_pert} up to first order to calculate the frequency shift upon variation of the unit cell expansion factor as shown in Figure \ref{fig:4}d of the main text.

\begin{thebibliography}{10}

\bibitem{joannopoulosPhotonicCrystalsMolding2008}
J.~D. Joannopoulos, editor,
 {\em Photonic Crystals: Molding the Flow of Light}, 2nd ed ed.
  ({Princeton University Press}, {Princeton}, 2008).

\bibitem{pitruzzelloPhotonicCrystalResonances2018}
G.~Pitruzzello and T.~F. Krauss,
 J. Opt. {\bf 20}, 073004 (2018).

\bibitem{hafeziRobustOpticalDelay2011a}
M.~Hafezi, E.~A. Demler, M.~D. Lukin, and J.~M. Taylor,
 Nature Phys {\bf 7}, 907 (2011), arXiv:1102.3256.

\bibitem{dietrichGaAsIntegratedQuantum2016}
C.~P. Dietrich, A.~Fiore, M.~G. Thompson, M.~Kamp, and S.~H{\"o}fling,
 Laser \& Photonics Reviews {\bf 10}, 870 (2016), arXiv:1601.06956.

\bibitem{srinivasanMomentumSpaceDesign2002}
K.~Srinivasan and O.~Painter,
 Opt. Express, OE {\bf 10}, 670 (2002).

\bibitem{akahaneHighPhotonicNanocavity2003}
Y.~Akahane, T.~Asano, B.-S. Song, and S.~Noda,
 Nature {\bf 425}, 944 (2003).

\bibitem{akahaneFinetunedHighQPhotoniccrystal2005}
Y.~Akahane, T.~Asano, B.-S. Song, and S.~Noda,
 Opt. Express, OE {\bf 13}, 1202 (2005).

\bibitem{foremanWhisperingGalleryMode2015}
M.~R. Foreman, J.~D. Swaim, and F.~Vollmer,
 Adv. Opt. Photon., AOP {\bf 7}, 168 (2015).

\bibitem{hasan_colloquium_2010}
M.~Z. Hasan and C.~L. Kane,
 Rev. Mod. Phys. {\bf 82}, 3045 (2010), arXiv:1002.3895.

\bibitem{RevModPhys.83.1057}
X.-L. Qi and S.-C. Zhang,
 Rev. Mod. Phys. {\bf 83}, 1057 (2011), arXiv:1008.2026.

\bibitem{asboth2016short}
J.~K. Asb{\'o}th, L.~Oroszl{\'a}ny, and A.~P{\'a}lyi,
 Lecture notes in physics {\bf 919}, 166 (2016), arXiv:1509.02295.

\bibitem{renTopologicalPhasesTwodimensional2016a}
Y.~Ren, Z.~Qiao, and Q.~Niu,
 Rep. Prog. Phys. {\bf 79}, 066501 (2016), arXiv:1612.09127.

\bibitem{haldanePossibleRealizationDirectional2008}
F.~D.~M. Haldane and S.~Raghu,
 Phys. Rev. Lett. {\bf 100}, 013904 (2008), arXiv:cond-mat/0503588.

\bibitem{fuTopologicalCrystallineInsulators2011}
L.~Fu,
 Phys. Rev. Lett. {\bf 106}, 106802 (2011), arXiv:1010.1802.

\bibitem{PhysRevLett.114.223901}
L.-H. Wu and X.~Hu,
 Phys. Rev. Lett. {\bf 114}, 223901 (2015), arXiv:1509.00919.

\bibitem{Barik_2016}
S.~Barik, H.~Miyake, W.~DeGottardi, E.~Waks, and M.~Hafezi,
 New Journal of Physics {\bf 18}, 113013 (2016), arXiv:1605.08822.

\bibitem{andersonUnidirectionalEdgeStates2017}
P.~D. Anderson and G.~Subramania,
 Opt. Express {\bf 25}, 23293 (2017).

\bibitem{barikTopologicalQuantumOptics2018}
S.~Barik \emph{et\,\,al.},
 Science {\bf 359}, 666 (2018), arXiv:1711.00478.

\bibitem{aroraDirectQuantificationTopological2021}
S.~Arora, T.~Bauer, R.~Barczyk, E.~Verhagen, and L.~Kuipers,
 Light: Science \& Applications {\bf 10}, 9 (2021), arXiv:2008.06497.

\bibitem{maAllSiValleyHallPhotonic2016}
T.~Ma and G.~Shvets,
 New J. Phys. {\bf 18}, 025012 (2016), arXiv:1601.06673.

\bibitem{dongValleyPhotonicCrystals2017}
J.-W. Dong, X.-D. Chen, H.~Zhu, Y.~Wang, and X.~Zhang,
 Nature Mater {\bf 16}, 298 (2017), arXiv:1709.05190.

\bibitem{parappurathDirectObservationTopological2020}
N.~Parappurath, F.~Alpeggiani, L.~Kuipers, and E.~Verhagen,
 Science Advances {\bf 6}, eaaw4137 (2020), arXiv:1811.10739.

\bibitem{bliokhQuantumSpinHall2015}
K.~Y. Bliokh, D.~Smirnova, and F.~Nori,
 Science {\bf 348}, 1448 (2015), arXiv:1504.03113.

\bibitem{salaSpinOrbitCouplingPhotons2015}
V.~G. Sala \emph{et\,\,al.},
 Phys. Rev. X {\bf 5}, 011034 (2015), arXiv:1406.4816.

\bibitem{lodahlChiralQuantumOptics2017}
P.~Lodahl \emph{et\,\,al.},
 Nature {\bf 541}, 473 (2017), arXiv:1608.00446.

\bibitem{barikChiralQuantumOptics2020}
S.~Barik, A.~Karasahin, S.~Mittal, E.~Waks, and M.~Hafezi,
 Phys. Rev. B {\bf 101}, 205303 (2020), arXiv:1906.11263.

\bibitem{jiRobustFanoResonance2021}
C.-Y. Ji, Y.~Zhang, B.~Zou, and Y.~Yao,
 Phys. Rev. A {\bf 103}, 023512 (2021), arXiv:2005.10426.

\bibitem{kimMultibandPhotonicTopological2021}
K.-H. Kim and K.-K. Om,
 Advanced Optical Materials {\bf 9}, 2001865 (2021).

\bibitem{mehrabadChiralTopologicalPhotonics2020}
M.~J. Mehrabad \emph{et\,\,al.},
 Optica, OPTICA {\bf 7}, 1690 (2020), arXiv:1912.09943.

\bibitem{zengElectricallyPumpedTopological2020a}
Y.~Zeng \emph{et\,\,al.},
 Nature {\bf 578}, 246 (2020), arXiv:1905.03671.

\bibitem{sirokiTopologicalPhotonicsCrystals2017}
G.~Siroki, P.~A. Huidobro, and V.~Giannini,
 Phys. Rev. B {\bf 96}, 041408 (2017), arXiv:1703.09248.

\bibitem{yangTopologicalWhisperingGallery2018}
Y.~Yang and Z.~H. Hang,
 Opt. Express, OE {\bf 26}, 21235 (2018).

\bibitem{gaoDiracvortexTopologicalCavities2020}
X.~Gao \emph{et\,\,al.},
 Nature Nanotechnology , 1 (2020), arXiv:1911.09540.

\bibitem{jalalimehrabadSemiconductorTopologicalPhotonic2020}
M.~Jalali~Mehrabad \emph{et\,\,al.},
 Appl. Phys. Lett. {\bf 116}, 061102 (2020), arXiv:1910.07448.

\bibitem{sunTopologicalRingcavityLaser2021}
X.-C. Sun and X.~Hu,
 Phys. Rev. B {\bf 103}, 245305 (2021), arXiv:1906.02464.

\bibitem{hafeziImagingTopologicalEdge2013}
M.~Hafezi, S.~Mittal, J.~Fan, A.~Migdall, and J.~M. Taylor,
 Nature Photon {\bf 7}, 1001 (2013).

\bibitem{gorlachFarfieldProbingLeaky2018}
M.~A. Gorlach \emph{et\,\,al.},
 Nat Commun {\bf 9}, 909 (2018), arXiv:1705.04236.

\bibitem{jagerskaRefractiveIndexSensing2010}
J.~J{\'a}gersk{\'a}, H.~Zhang, Z.~Diao, N.~L. Thomas, and R.~Houdr{\'e},
 Opt. Lett., OL {\bf 35}, 2523 (2010).

\bibitem{hamelSpontaneousMirrorsymmetryBreaking2015}
P.~Hamel \emph{et\,\,al.},
 Nature Photon {\bf 9}, 311 (2015), arXiv:1411.6380.

\bibitem{iwahashiHigherorderVectorBeams2011}
S.~Iwahashi \emph{et\,\,al.},
 Opt. Express, OE {\bf 19}, 11963 (2011).

\bibitem{yangSpinMomentumLockedEdgeMode2020}
Z.-Q. Yang, Z.-K. Shao, H.-Z. Chen, X.-R. Mao, and R.-M. Ma,
 Phys. Rev. Lett. {\bf 125}, 013903 (2020).

\bibitem{shaoHighperformanceTopologicalBulk2020b}
Z.-K. Shao \emph{et\,\,al.},
 Nat. Nanotechnol. {\bf 15}, 67 (2020).

\bibitem{yuCriticalCouplingsTopologicalinsulator2021}
S.-Y. Yu \emph{et\,\,al.},
 National Science Review {\bf 8}, nwaa262 (2021), arXiv:2008.09547.

\bibitem{reardon_fabrication_2012}
C.~P. Reardon, I.~H. Rey, K.~Welna, L.~O'Faolain, and T.~F. Krauss,
 Journal of Visualized Experiments : JoVE  (2012).

\bibitem{portalupiPlanarPhotonicCrystal2010}
S.~L. Portalupi \emph{et\,\,al.},
 Opt. Express {\bf 18}, 16064 (2010).

\bibitem{orazbayevQuantitativeRobustnessAnalysis2019}
B.~Orazbayev and R.~Fleury,
 Nanophotonics {\bf 8}, 1433 (2019).

\bibitem{depazEngineeringFragileTopology2019}
M.~B. {de Paz}, M.~G. Vergniory, D.~Bercioux, A.~{Garc{\'i}a-Etxarri}, and
  B.~Bradlyn,
 Phys. Rev. Research {\bf 1}, 032005 (2019), arXiv:1903.02562.

\bibitem{palmerBerryBandsPseudospin2021}
S.~J. Palmer and V.~Giannini,
 Phys. Rev. Research {\bf 3}, L022013 (2021), arXiv:2009.07033.

\bibitem{hongBackgroundFreeDetectionSingle2011}
X.~Hong \emph{et\,\,al.},
 Nano Lett. {\bf 11}, 541 (2011).

\bibitem{rubinsztein-dunlopRoadmapStructuredLight2016}
H.~{Rubinsztein-Dunlop} \emph{et\,\,al.},
 J. Opt. {\bf 19}, 013001 (2016).

\bibitem{se-heon_kim_symmetry_2003}
S.-H. Kim and Y.-H. Lee,
 IEEE Journal of Quantum Electronics {\bf 39}, 1081 (2003).

\bibitem{luSelectiveEngineeringCavity2014}
X.~Lu, S.~Rogers, W.~C. Jiang, and Q.~Lin,
 Appl. Phys. Lett. {\bf 105}, 151104 (2014), arXiv:1407.4488.

\bibitem{johnsonPerturbationTheoryMaxwell2002}
S.~G. Johnson \emph{et\,\,al.},
 Phys. Rev. E {\bf 65}, 066611 (2002).

\bibitem{iwamotoRecentProgressTopological2021}
S.~Iwamoto, Y.~Ota, and Y.~Arakawa,
 Opt. Mater. Express, OME {\bf 11}, 319 (2021).

\bibitem{COMSOL52a}
{COMSOL AB},
 {\em COMSOL Multiphysics® v. 5.3a.} .

\end{thebibliography}

\begin{thebibliography}{1}
\providecommand{\url}[1]{\texttt{#1}}
\providecommand{\urlprefix}{URL }

\bibitem{joannopoulosPhotonicCrystalsMolding2008_SI}
J.~D. Joannopoulos, editor,
\newblock {\em Photonic Crystals: Molding the Flow of Light}, 2nd ed ed.
  ({Princeton University Press}, {Princeton}, 2008).

\bibitem{johnsonPerturbationTheoryMaxwell2002_SI}
S.~G. Johnson {\em et\,\,al.},
\newblock Phys. Rev. E {\bf 65}, 066611 (2002).

\end{thebibliography}
\end{document}